%% file: main.tex
\documentclass[conference]{IEEEtran}
\IEEEoverridecommandlockouts
\usepackage{cite}
\usepackage{amsmath,amssymb,amsfonts}
\usepackage{graphicx}
\usepackage{textcomp}
\usepackage{xcolor}
\def\BibTeX{{\rm B\kern-.05em{\sc i\kern-.025em b}\kern-.08em
    T\kern-.1667em\lower.7ex\hbox{E}\kern-.125emX}}

\usepackage{subfigure}
\usepackage[noend]{algpseudocode}
\usepackage{enumitem}
\usepackage{booktabs} 
\usepackage{balance}
\usepackage{algorithm}
\usepackage{subfigure}
\usepackage[noend]{algpseudocode}
\usepackage{enumitem}
\usepackage{hyperref}
\usepackage{comment}
\usepackage{bm}

\begin{document}


\title{Mitigating Frequency Bias in Next-Basket Recommendation via Deconfounders}



\author{
    \IEEEauthorblockN{Xiaohan Li\IEEEauthorrefmark{1}\IEEEauthorrefmark{2}, Zheng Liu\IEEEauthorrefmark{1}\IEEEauthorrefmark{3}, Luyi Ma\IEEEauthorrefmark{2}, Kaushiki Nag\IEEEauthorrefmark{2}, Stephen Guo\IEEEauthorrefmark{4}, Philip S. Yu\IEEEauthorrefmark{3}, Kannan Achan\IEEEauthorrefmark{2}}
    \IEEEauthorblockA{\IEEEauthorrefmark{2}Walmart Global Tech, Sunnyvale, CA, USA
    \\\{xiaohan.li, luyi.ma, kaushiki.nag, stephen.guo, kannan.achan\}@walmart.com}
    \IEEEauthorblockA{\IEEEauthorrefmark{3}University of Illinois at Chicago, Chicago, IL, USA
    \\\{zliu212, psyu\}@uic.edu}
    \IEEEauthorblockA{\IEEEauthorrefmark{4}Indeed, CA, USA\\sguo@indeed.com}
    \thanks{\IEEEauthorrefmark{1}Both authors contributed equally to this research.}
}

\maketitle

\input{text/1-Abstract}





\input{text/2-Introduction}

\input{text/2.1_Frequency_Bias}

\input{text/3-Method}
\input{text/4-Experiments}

\input{text/5-RelatedWorks}

\input{text/6-Conclusion}

\bibliographystyle{IEEEtran}
\balance
\bibliography{ref} 

\end{document}

%% file: text/1-Abstract.tex
\begin{abstract}
Recent studies on Next-basket Recommendation (NBR) have achieved much progress by leveraging Personalized Item Frequency (PIF) as one of the main features, which measures the frequency of the user's interactions with the item. However, taking the PIF as an explicit feature incurs bias towards frequent items. Items that a user purchases frequently are assigned higher weights in PIF-based recommender system and  appear more frequently in the personalized
recommendation list. 
As a result, the system will lose the fairness and balance between items that the user frequently purchases and items that the user never purchases.
We refer to this systematic bias on personalized recommendation lists
as frequency bias, which narrows users' browsing scope and reduces the system utility.
We adopt causal inference theory to address this issue. Considering the influence of historical purchases on users' future interests, the user and item representations can be viewed as unobserved confounders in the causal diagram. In this paper, we propose a deconfounder model named \textbf{FENDER }(\underline{F}r\underline{e}que\underline{n}cy-aware \underline{D}econfounder for N\underline{e}xt-basket \underline{R}ecommendation) to mitigate the frequency bias. 
With the deconfounder theory and the causal diagram we propose, FENDER decomposes PIF with a neural tensor layer to obtain substitute confounders for users and items. Then, FENDER performs unbiased recommendations considering the effect of these substitute confounders.
Experimental results demonstrate that FENDER has derived diverse and fair results compared to ten baseline models on three datasets while achieving competitive performance. Further experiments illustrate how FENDER balances users' historical purchases and potential interests.

\end{abstract}

%% file: text/2-Introduction.tex
\section{Introduction}

Next-basket Recommendation (NBR) \cite{yu2016dynamic, le2017basket, wan2018representing, liu2020basket, hu2020modeling} means to recommend a list of items (the next basket) that a user may be interested using the content of previous baskets. 
Unlike other
recommendation scenario, a unique characteristics of NBR is its \textit{repetitive purchase pattern}, which means the same item may appear multiple times in baskets that are close together in time. For example,  a user may purchase his favorite potato chips frequently during a short period of time.
In contrast, user usually does not interact with the same item very often in other recommendation scenarios, such as movie and book recommendation. 


To capture the pattern of repetitive purchase in NBR,
previous models strongly rely on \textbf{Personalized Item Frequency (PIF)} in their decision-making processes
\cite{hu2020modeling}. 
PIF denotes the appearance times of the item in the user's purchased baskets divided by the number of baskets. For example, Mary has made five purchased baskets on the platform, three containing toilet paper. Then, the PIF of toilet paper with respect to Mary is 0.6. PIF measures the frequency of the user's interactions with the product, and the higher PIF is, the more likely the user would buy it next time.
\cite{hu2020modeling, faggioli2020recency, hu2019sets2sets} claim that a high PIF of an item to a user indicates the user's high preference for the item.
Hu et al. \cite{hu2020modeling} propose a KNN-based model regarding PIF as the main feature which achieves the state-of-the-art performance on NBR.

However, the reliance on PIF can be pernicious, especially when the weight of PIF in the model is high. Suppose a recommendation model relies solely on PIF to make decisions. It will only recommend items the user has purchased before. In this case, though the model can achieve high performance due to repetitive purchase pattern, it does not derive any new knowledge for users.
Based on this example, we conclude that the improper use of PIF can cause systematic bias in NBR. Here we illustrate where the systematic bias comes from. PIF-based methods assume that a higher PIF indicates a higher preference of a user for an item and leads to a higher probability of the next purchase. Thus the model takes PIF as a crucial feature in prediction. However, high PIF, which means the existence of previous purchases, has reversely been caused by the user's preference before. Therefore, if the model simply takes PIF as a feature, the model will erroneously take historical information directly as predictive output, reinforcing the system's bias and users' stereotype. In this paper, we name this systematic bias caused by the misuse of PIF \textbf{frequency bias}. Frequency bias can also be viewed as ``personalized popularity bias", expressing a false tendency of RS to recommend previous purchased items of the user too often.

Many studies have discussed the unbiasedness or fairness issue on recommender systems recently \cite{schnabel2016recommendations,wang2020information,liu2020general,wang2020causal,wang2021deconfounded, wang2022unbiased}.
Although most of these approaches are not designed for NBR and inapplicable to our problem directly, they inspired us to address frequency bias from a causal inference perspective. Here we illustrate the causal diagram we use.
First, we claim that PIF can directly affect a user's next-basket purchase due to some mechanism such as Mere Exposure Effect \cite{bornstein1992stimulus} or feedback loop~\cite{chaney2018algorithmic}. 
Moreover, we argue that many other factors of users or items, such as user's personality and item's characteristic, can affect both the PIF and the next-basket purchase simultaneously, and thus are confounders in the causal diagram. 
For example, suppose the user's personality (adventurous or prudent) is a confounder. Since it can affect both the PIF (cause) and the next-basket purchase (effect). An adventurous user would like to try new items that (s)he hasn't bought before, but a prudent user would only purchase what (s)he has bought before. 
Therefore, the causality between PIF and next-basket purchase depends on the user's personality and cannot be identified without observing the confounder.


In this paper, we propose a novel framework named  \textbf{FENDER}  (\underline{F}r\underline{e}que\underline{n}cy-aware \underline{D}econfounder for N\underline{e}xt-basket  \underline{R}ecommendation) based on the deconfounder theory to mitigate the frequency bias in NBR.
Inspired by the deconfounder \cite{wang2019blessings, liu2022mitigating} theory, FENDER addresses the unobserved confounders of PIF by leveraging a two-stage training schema.
In the first stage, we use a latent factor model (i.e., Neural Tensor Layer \cite{he2017neural,socher2013reasoning}) to model the PIF for each user-item pair. In the second stage, we use embeddings learned from the first model to predict the next basket, and jointly learn user and item embeddings. 
In the inference step, we can adopt flexible weight to balance the influence of PIF on the next-basket purchase. 
We can also adjust FENDER's recommendation style from conservative (prefer high PIF items) to diverse (prefer low PIF items) by adjusting the weight in the inference step.

In summary, here are the contributions of this paper:
\begin{itemize}
    \item To the best of our knowledge, this is the first study that analyzes the repetitive purchase pattern of NBR from the causal inference perspective. We propose the concept \textit{frequency bias}, describing the preferences of the products having been purchased by users in the personalized recommender system.
    \item We propose FENDER, a powerful, flexible, and robust two-stage framework to model and mitigate the frequency bias by adopting the deconfounder, which models the historical preference and the next-basket purchase, respectively. By calculating the personalized penalty of each user-item pair, FENDER can derive unbiased next-basket recommendations varying from conservative to diverse styles.
    \item Comprehensive analysis and experiments are conducted to support our propositions from three aspects. First, we compare FENDER with ten baseline models to show its superiority in NBR. Second, we propose a new metric named \textit{negative Top-Frequency Ratio} (nTFR) to measure the diversity in the predicted baskets. A case study is conducted to further illustrate the diverse recommendations of FENDER. The third experiment demonstrates the robustness of FENDER when dealing with manually interfered data.
\end{itemize}

%% file: text/2.1_Frequency_Bias.tex
\section{Backgrounds}
\begin{figure}[ht]
\centering
    \includegraphics[height=1.7in]{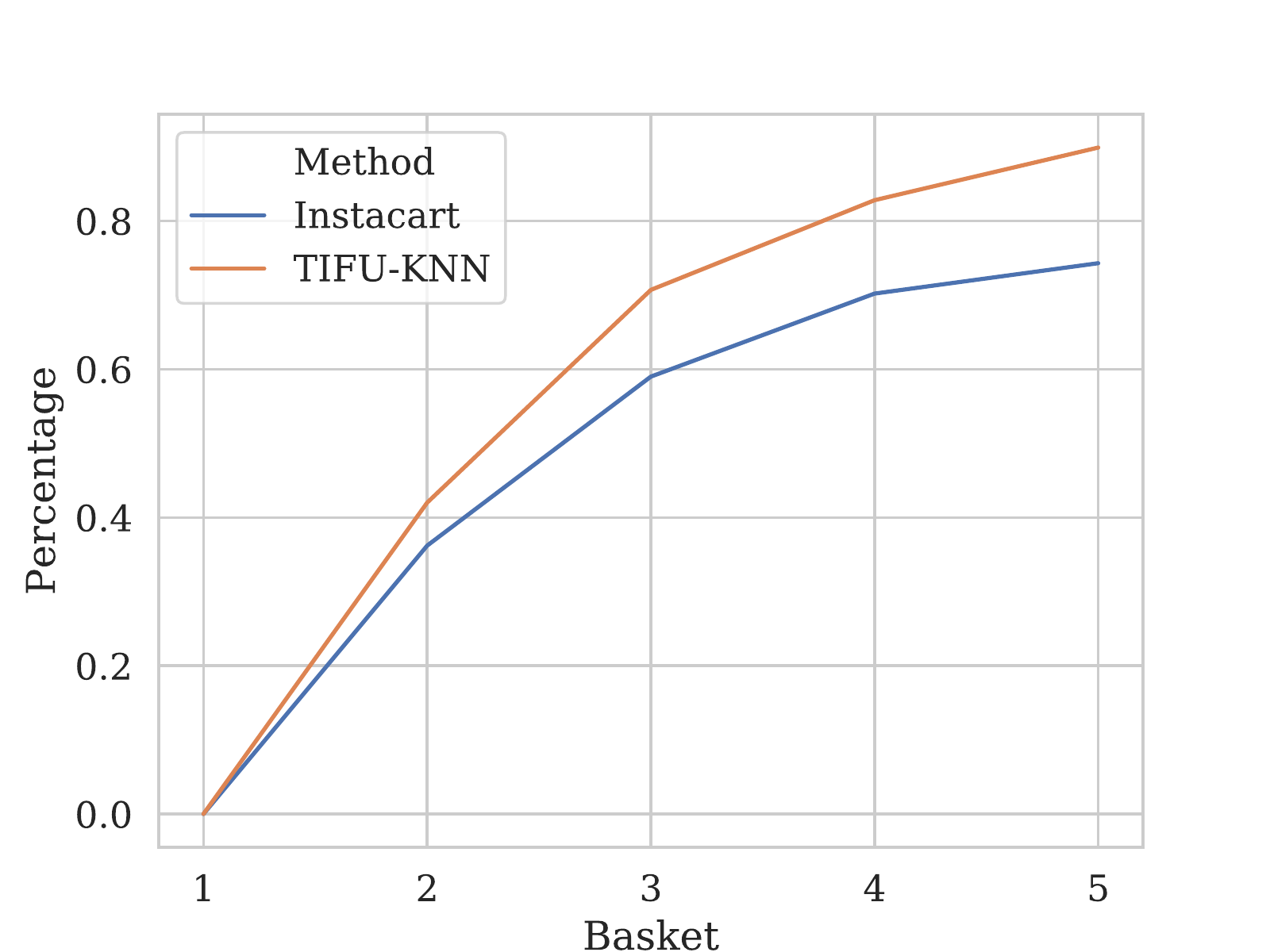}
    \caption{The percentage of items in each basket that have also been purchased in the previous baskets.} 
    \label{fig:pif}
\end{figure}
\subsection{Frequency Bias}

In this section, we demonstrate the manifestation of frequency bias. Frequency bias describes the bias of an NBR model on items purchased by the user before. Figure \ref{fig:pif} shows the percentage of items that have been purchased in previous baskets based on the Instacart Dataset \cite{instacart} and the prediction results of the state-of-the-art NBR model TIFU-KNN \cite{hu2020modeling}. To obtain this figure, we select users with more than five baskets and calculate the ratio of items existing in previous baskets. For example, the percentage is zero for the first basket since there is no previous basket. Next, for the second basket, we calculate 
how many items in it are also in the first basket. Then, for the third basket, we calculate the percentage of items in it that are also in the first and second baskets. We plot the blue line in Figure \ref{fig:pif} to show this for each basket in the Instacart dataset. To obtain the orange line in the figure, we adopt the results of TIFU-KNN to substitute the Instacart data in the blue line. We train a recommendation model for each basket based on its previous baskets and calculate this percentage based on the recommendation results.

From Figure \ref{fig:pif}, we observe that the result of TIFU-KNN (orange line) is always higher than the ground-truth data (blue line), especially when the basket number is high. This observation also fits our intuition: recommendation algorithms tend to provide a higher ranking to items the user has purchased before. These items are also more likely to be included in the next-basket recommendation. However, though these algorithms may achieve good performance on the first several baskets, they are biased since their prediction distribution differs from the ground truth. Furthermore, these methods will deteriorate the long-term benefits of the accuracy and diversity of the recommendation results. 


\subsection{Fairness and Debiasing on Recommender Systems}

As a crucial part of large-scale online platforms, the fairness and debiasing issue on Recommender Systems is essential to improve the user experience and system utility. However, conventional recommendation algorithms mainly focus on improving prediction accuracy, especially improving evaluation metrics' value. Most studies do not dissect the recommendation results they derive and thus neglect the difference between their recommendation results and the ground truth in data distribution.

Recently, many studies have focused on mitigating systematic biases and maintaining fairness of Recommender Systems. According to their literature, many factors, such as feedback loop \cite{chaney2018algorithmic, wang2021deconfounded}, missing-not-at-random nature \cite{yang2018unbiased,saito2020unbiased}, and exposure/popularity/position biases \cite{liang2016modeling,zhang2021causal,joachims2017unbiased}, can affect the observational collaborative information and further deteriorate the whole system. Though focusing on different problems, these studies inspired our work dealing with frequency bias.

In this paper, we have introduced the frequency bias and its outcomes in NBR, and will address this issue via causal inference theories. From the causal inference perspective \cite{pearl2009causality,rubin2005causal,wang2020causal}, recommendation is naturally a causal inference problem with unobserved factors. A purchase is a combined result of the user's preferences together with item characteristics, while both the user's preferences and item characteristics are unobservable to us. In other to deal with these latent factors, we adopt the deconfounder framework to address frequency bias.

Wang and Blei \cite{wang2019blessings,wang2020towards} propose the deconfounder theory to deal with the effect of multiple causes. A deconfounder is constructed based on the belief that the distribution of multiple causes contains information about the confounders. Hence, it resorts to a factor model to capture the causes' distribution and use this model to correct the confounding bias.

%% file: text/3-Method.tex
\section{Preliminaries}

\subsection{Notations \& Problem Definition}
Consider there is a set of user $U$ and a set of item $I$ in NBR and each user consists of $T$ baskets
 $\mathcal{B}_{T}^{u} = \{ \bm B_1^u, \bm B_2^u, \cdots, 
\bm B_{T}^u \}$. 
Each basket $\bm B$ contains a set of items $\{ i_1, i_2,\cdots,i_{|\bm B|}\}$ without duplicates, denoting the items that the user purchased in this basket.
In this paper, we use letter $r$ to denote the \textbf{probability of purchase} in the dataset. $r_{(u,i,t)} = 1$ if the user $u$ purchase item $i$ in the $t$-th basket and vice versa.

The goal of next-basket recommendation can be viewed as the prediction of next basket $\bm B_{t}^u$ with $\mathcal{B}_{t-1}^u$ given. In other words, the model needs to predict $\hat{r}_{(u,i,t)}$ for user $u$ and item $i$ using the records of previous $t-1$ baskets. 

\paragraph{Personalized Item Frequency (PIF)}
As a crucial feature in recent next-basket recommendation models~\cite{hu2020modeling, faggioli2020recency}, PIF is the number of times that the item was purchased by the user divided by the number of baskets. Formally, for each pair of user $u$ and item $i$, the PIF at the $t$-th basket can be defined as 

\begin{equation}
    PIF_{(u,i,t)} =\frac{  \sum_{t'=1}^{t-1} r_{(u,i,t')}  } {t-1}.
\end{equation}

For example, a user $u$ has made three purchases at a grocery store, and two of those records contain milk. Then, the PIF of milk at the fourth basket is $p_{(u,`milk',4)}= 2/3$. It is obvious that the PIF of each $(u,i,t)$ triplet can be easily calculated with linear complexity. 
Intuitively, a higher PIF indicates a higher preference of the product and can lead to higher probability of purchase in the next basket.

\subsection{Basic Causal Graphs in NBR}

Given historical purchase records, one can build effective next-basket recommender systems based on various causal diagrams and each causal diagram is a unique assumption of data latent structure. Here we first introduce two straightforward recommendation methods together with their corresponding causal graphs to show the connection between the recommendation algorithm and the corresponding causal diagram.

\subsubsection{Simple PIF-based Recommendation}

A straightforward assumption of PIF is that the higher the PIF is, the more likely the user is to purchase the item in the next basket.
This assumption is related to the causal diagram Figure \ref{fig:tra}, meaning that the probability of a purchase merely depends on the frequency of historical purchases. 
Therefore, we can build a recommender system based on this causal diagram, merely ranking the PIF for each item:

\begin{equation}
    \hat{r}_{(u,i,t)} =   PIF_{(u,i,t)}.
\end{equation}

Such a system is easy to implement and can even achieve comparable results in offline evaluation \cite{hu2020modeling}. However, since it never recommends new items for users, this system is actually with low utility and deteriorates users' experience.

\subsubsection{Conventional Recommender Systems}

\begin{figure}[tbp]
\centering
\subfigure[Simple PIF-based recommendation.]{
\vspace{0.25in}
\includegraphics[height=0.25in]{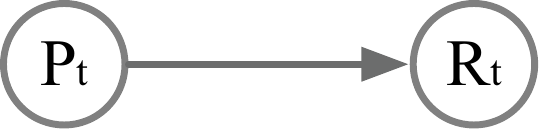}	
\vspace{0.25in}
\label{fig:tra}
}
\quad
\hspace{0.005\textwidth}
\subfigure[Conventional recommender systems.]{
\includegraphics[height=1in]{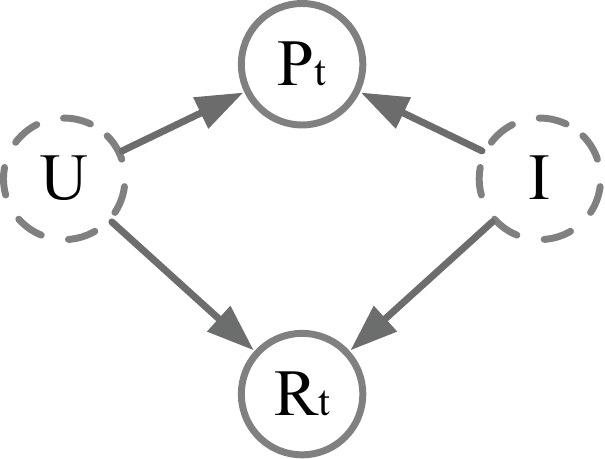}
\label{fig:simple}
}

\subfigure[Frequency Bias.]{
\includegraphics[height=1.1in]{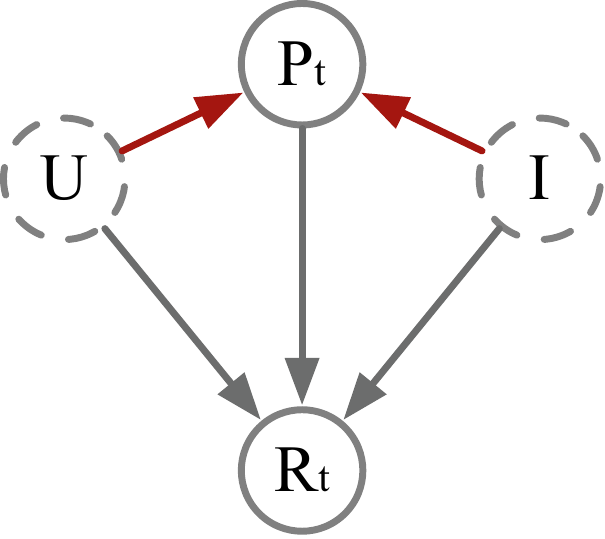}
\label{fig:groundtruth}
}
\quad
\hspace{0.005\textwidth}
\subfigure[FENDER based on deconfounder.]{
\includegraphics[height=1.2in]{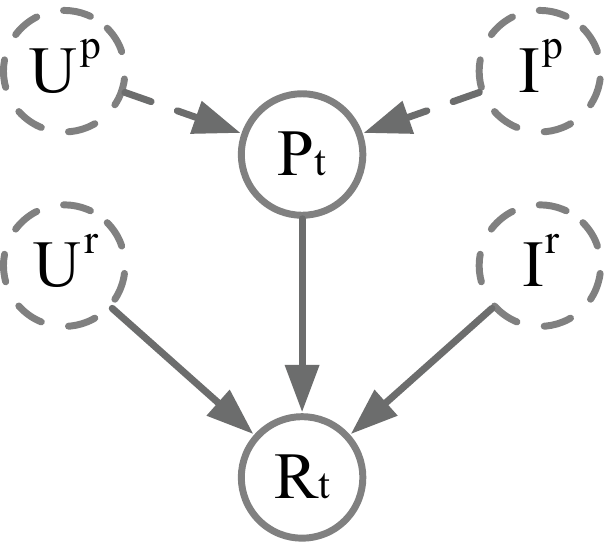}
\label{fig:deconfounder}
}
\caption{Causal graphs depicting causalities between User $U$, item $I$, PIF $P_{t}$ and purchase $R_t$. Dotted circles represent unobserved variables.}  
\label{fig:causal_graph}
\end{figure}

Most conventional recommendation methods (e.g., Matrix Factorization (MF) \cite{rendle2012bpr}) are based on the causal diagram shown in Figure \ref{fig:simple}. In this diagram, there are unobserved user and item nodes that can represent their features. Both the probability of purchase at the $t$-th basket (denoted by $R_t$) and the PIF at the $t$-th basket (denoted by $P_{t}$) are both sampled based on the interaction between user and item nodes. Here we take MF as an example \cite{mnih2007probabilistic}:

\begin{equation}
    \mathcal{L}_{BPR} = \sum_{(u,i,j)\in O_t} -\log \sigma(
    \langle \bm e_{u} \cdot \bm e_i \rangle - \langle \bm e_{u} \cdot \bm e_j \rangle ) +\lambda_{MF} \left\Vert \Theta_{MF} \right\Vert_2.
    \label{bpr}
\end{equation}

Above is the loss function to deploy MF with BPR loss and $L^2$ regularizor $\left\Vert \Theta_{MF} \right\Vert_2$ to predict next basket at timestep $t$.
In this function, $O_t = \{(u,i,j)| PIF_{(u,i,t)} > 0,  PIF_{(u,j,t)} = 0 \} $ is a set of $(u,i,j)$ triplet where $u$ is a user and $i$, $j$ are positive and negative samples, respectively. Formally, 
In this objective, the probability of purchase $\hat{r}_{(u,i,t)} = \langle \bm e_{u} \cdot \bm e_i \rangle$ is represented as the inner product of user embedding $\bm e_u$
and item embedding $\bm e_i$. In this case, the user and item nodes in the causal diagram represents the corresponding embedding vectors in the BPR-MF model. Both of above two methods can derive predictions on NBR.
However,
our experiments on TIFU-KNN and BPR-MF demonstrate that they do not perform very well in fairness, indicating their causal diagrams may be defective.

\section{Method}
In this paper, we propose a novel framework FENDER based on deconfounder theory to address frequency bias in NBR. 

\subsection{Causal Graph}
The causal diagram in Figure \ref{fig:groundtruth} illustrates the cause of frequency bias. Most NBR model assumes an intuitive assumption: higher PIF, representing a higher purchase frequency, can lead to a higher probability of future purchases. However, we argue that this assumption is not comprehensive. On the one hand, a user's purchase behavior is caused by the user and item attributes; on the other hand, these attributes can also lead to high PIFs, since the user's interest and item's characteristics are not likely to change quickly. Moreover, some unobserved mechanisms, such as the Mere Exposure Effect \cite{bornstein1992stimulus}, and the feedback loop~\cite{chaney2018algorithmic} lead to a direct causal relationship between high PIF and users' purchases. To describe the causality between these factors, the causal diagram in Figure \ref{fig:groundtruth} adopts four nodes in total: the user's interest (denoted by \raisebox{.3pt}{\textcircled{\raisebox{-.9pt} {$U$}}}  ), the item's characteristics (denoted by \raisebox{.3pt}{\textcircled{\raisebox{-.9pt} {$I$}}}), PIF indicating historical purchases (denoted by \raisebox{.3pt}{\textcircled{\raisebox{-.9pt} {$P_{t}$}}}) and the next-basket purchase (denoted by \raisebox{.3pt}{\textcircled{\raisebox{-.9pt} {$R_{t}$}}}). In this graph, the user node \raisebox{.3pt}{\textcircled{\raisebox{-.9pt} {$U$}}} and item node \raisebox{.3pt}{\textcircled{\raisebox{-.9pt} {$I$}}} point to \raisebox{.3pt}{\textcircled{\raisebox{-.9pt} {$P_{t}$}}} and \raisebox{.3pt}{\textcircled{\raisebox{-.9pt} {$R_{t}$}}} simultaneously, which means the user and item's characteristics affect both historical and next-basket purchases. Moreover, a direct arrow from PIF \raisebox{.3pt}{\textcircled{\raisebox{-.9pt} {$P_{t}$}}} to next-basket purchase \raisebox{.3pt}{\textcircled{\raisebox{-.9pt} {$R_{t}$}}} represents the direct influence of previous user choice on the next-basket purchase.

From this causal diagram, we can observe the frequency bias comes from the confounding bias between \raisebox{.3pt}{\textcircled{\raisebox{-.9pt} {$P_{t}$}}} and \raisebox{.3pt}{\textcircled{\raisebox{-.9pt} {$R_{t}$}}}. In causal inference theory, confounders are factors that can affect both the cause and effect. In this case, \raisebox{.3pt}{\textcircled{\raisebox{-.9pt} {$U$}}} and \raisebox{.3pt}{\textcircled{\raisebox{-.9pt} {$I$}}} are unobserved confounders that affect both the cause \raisebox{.3pt}{\textcircled{\raisebox{-.9pt} {$P_{t}$}}} and the effect \raisebox{.3pt}{\textcircled{\raisebox{-.9pt} {$R_{t}$}}}, which cause confounding bias. Confounders in the causal graph can affect both the cause and effect and further control their association. 
In our case, for example, suppose the user’s personality can be categorized into adventurous or prudent. Then, an adventurous user prefers to try new items that (s)he hasn’t bought before, and the association between \raisebox{.3pt}{\textcircled{\raisebox{-.9pt} {$P_{t}$}}} and \raisebox{.3pt}{\textcircled{\raisebox{-.9pt} {$R_{t}$}}} for (s)he is lower. Reversely, the association between them for a prudent user can be higher. Therefore, without addressing the confounding issue, even if we measure the correlation between \raisebox{.3pt}{\textcircled{\raisebox{-.9pt} {$P_{t}$}}} and \raisebox{.3pt}{\textcircled{\raisebox{-.9pt} {$R_{t}$}}} from the observational dataset, it can represent neither group of users and is unreliable when data distribution shifts.

\subsection{Deconfounder}

In order to deal with the unobserved confounders, our method is motivated by the theory of deconfounder~\cite{wang2019blessings,wang2020towards}. 
Here we list the assumptions that the deconfounder requires in the following.

\paragraph{\textbf{Stable unit treatment value assumption (SUTVA)}}
The SUTVA assumption assumes that the potential outcomes for a particular unit only depend on the treatment to which the unit itself was received~\cite{rubin1980randomization,rubin2005causal}. In NBR, it means whether a user buys an item only related to the user and the item themselves, and is irrelevant to all other users or items.

\paragraph{\textbf{Overlap (Positivity)}}
Overlap asserts that, given the unobserved confounder, the conditional probability of any possible value of treatments is positive. In NBR, this means we assume
\begin{equation}
    0 < r_{(u,i,t)} < 1, \forall u,i,t,
\end{equation}
which means the probability of a user purchasing any item is ranging from 0 to 1.

\paragraph{\textbf{Single ignorability}}
Unlike the traditional unconfoundedness assumption requires ``All confounders observed'', the deconfounder requires ``No unobserved single-cause confounders'' assumption \cite{wang2019blessings}, which means we observe any confounders that affect only one of the causes. In NBR, apart from the collaborative information, the model does not have to observe any user or item features. Therefore, we assume that if a factor (of user $u$ or item $i$) can affect the purchase of $u$ on $i$, it must either affect $u$ purchasing other items or affect other users purchasing item $i$.

\subsection{FENDER}
Based on the three assumptions, we propose an novel unbiased framework  FENDER to mitigate the frequency bias in NBR. 
FENDER adopts a two-stage architecture to model the frequency bias before making a recommendation.

In the first step, FENDER models frequency bias by factorizing PIF. It learns the user and item embeddings $\bm e_{u,t}^{p}$, $\bm e_{i}^{p}$ with MF model with a neural tensor layer to factorize PIF, which is to learn not only the feature PIF itself, but also derive new knowledge about latent features that cause the PIF. In the second step, FENDER predicts the probability of items in the next basket. The learned embeddings of the first step are used to balance the current state and the previous state. We disentangle the user's previous interest $\bm e_{u,t}^{p}$ and current interest $\bm e_{u,t}^{r}$, and use them to derive unbiased recommendations in the inference step.

\subsubsection{First Stage Model}
In the first stage of the model, FENDER models the previous baskets by factorizing PIF at time step $t$. Since $PIF_{(u,i,t)}$ is calculated based on all previous baskets, it contains information about the historical preference of an item and can be used to correct frequency bias.
For each triplet in $\mathcal{I} =  \{(u,i,t)|u \in U, i \in I, 1 < t \leq T\}$, we can calculate $PIF_{(u,i,t)}$ from the original dataset $\{\mathcal{B}_u\}$.
In the first stage model, we decompose the PIF into two latent user and item representations. Motivated by \cite{he2017neural, socher2013reasoning}, a Neural Tensor Layer (NTL) is applied to generate the predicted PIF value from the latent embeddings. With Mean Squared Error (MSE) loss, the distance between the predicted PIF and the actual PIF will be minimized. Specifically, the process is as follows:

\begin{align}
    &p_{(u,i,t)} = \sigma \Bigg( \mathbf{h}^\intercal (\bm e_{u,t}^{p\intercal} \mathbf{W}_1^{[1:k]} \bm e_{i}^{p}) + \mathbf{W}_2  \begin{pmatrix}
         \bm e_{u,t}^{p} \\
         \bm e_{i}^{p}
    \end{pmatrix} + \mathbf{b}\Bigg), \\
    &\mathcal{L}_{1} = \frac{1}{ |\mathcal{I}| }   \sum_{ (u,i,t) \in \mathcal{I}} (p_{(u,i,t)} - PIF_{(u,i,t)})^2 +\lambda_{ {p}} \left\Vert \Theta_{ {p}} \right\Vert_2,
    \label{s1}
\end{align}

where $\bm e_{u,t}^{p}$ denoted the user embedding for user $u$ at the $t$-th basket, and $\bm e_{i}^{p}$ is the item embedding for item $i$ which is consistent to time. $\Theta_{p} = \{\bm W_1, \bm W_2, \bm b, \bm h\}$ is the parameter set of this model and $\lambda_{p}$ is its weight. 
$\mathcal{I} =  \{(u,i,t)|u \in U, i \in I, 1 < t \leq T\}$ denotes the sample set, and
$\mathbf{W}_1^{[1:k]} \in \mathbb{R}^{d \times d \times k}$ and $\mathbf{W}_2 \in \mathbb{R}^{k \times 2d}$ are 3d and 2d weight tensors, respectively.  These embeddings represent the feature PIF as well as other latent features that cause the PIF. The latent features help us know extra knowledge in addition to PIF to mitigate the frequency bias. In the next step, we will use this information to correct frequency bias when predicting the current basket purchase.

\subsubsection{Second Stage Model}
After we obtain $\bm e_{u,t}^{p}$ and $\bm e_{i}^{p}$ from the PIF decomposition stage, they are used in the second stage to predict the probability of next-basket purchase. The rationale of the second stage model is to balance the PIF and confounders in the model. In the second stage model, we learn another pair of user and item embeddings $\bm e_{u,t}^{r}$ and $\bm e_{i}^{r}$ to represent confounders. Here is the forward propagation formula to predict the probability of purchase $\hat{r}_{(u,i,t)}$ based on the causal graph shown in Figure \ref{fig:deconfounder}:

\begin{align}
    &p_{(u,i,t)} = \text{NTL}(\bm e_{u,t}^{p}, \bm e_{i}^{p}), \\
    &c_{(u,i,t)} = \text{NTL}(\bm e_{u,t}^{r}, \bm e_{i}^{r}), \\
    &\hat{r}_{(u,i,t)} = \omega \cdot c_{(u,i,t)} + (1-\omega) \cdot p_{(u,i,t)},
\end{align}

where $\text{NTL}$ is the Neural Tensor Layer and $\omega$ is the weight of the first factor. $p_{(u,i,t)}$ and $c_{(u,i,t)}$ are the embeddings of PIF and confounders. $\omega$ illustrates the inclination of PIF and confounders in the model training. In this formula, $\bm e_{u,t}^{p}$ and $\bm e_{i}^{p}$ stay fixed while other parameters and embeddings are trained with BPR \cite{rendle2012bpr} loss, which is 


\begin{equation}
    \mathcal{L}_{2} = \sum_{ (u,i,j)\in O_t } -\log\, \sigma(\hat{r}_{(u,i,t)} - \hat{r}_{(u,j,t)}) +\lambda_{ {r}} \left\Vert \Theta_{ {r}} \right\Vert_2,
    \label{bpr}
\end{equation}

where $O_t = \{ (u,i,j)| r_{(u,i,t)}=1, r_{(u,j,t)}=0\}$ is a set containing positive and negative items at the $t$-th basket. $\Theta_{r}$ is the parameter set of the second stage. The positive items are the items purchased in the $t$-th basket and we take it as the ground truth to train our model. In the inference step, we can easily change the recommendation style by tuning the value of weight $\omega$. The higher $\omega$ we set, the more FENDER relies on PIF to make recommendation, and thus there will be more repeat items in the recommendation list.



%% file: text/4-Experiments.tex
\section{Experiments}
This section evaluates our proposed FENDER on real and manually interfered datasets. The experimental results demonstrate that FENDER can better model user behavior and have more accurate and diverse recommendation results.

\subsection{Experiment Settings}
\subsubsection{Datasets}
We use three real-world next-basket recommendation datasets to measure FENDER's performance.
\begin{itemize}
    \item \textbf{\textit{Instacart \footnote{https://www.kaggle.com/c/instacart-market-basket-analysis}}}. This dataset was published by \textit{instacart.com}, an online grocery delivery and pick-up service platform in North America. It contains over 3 million orders and 200 thousand users. 
    
    \item \textbf{\textit{Dunnhumby \footnote{https://www.dunnhumby.com/source-files/}}}. We use the \textit{Let’s Get Sort-of-Real} dataset from \textit{Dunnhumby.com}. It is real in-store data, including the basket records for each user. 
    
    \item \textbf{\textit{Walmart Grocery \footnote{walmart.com}}}. This dataset is from Walmart Inc, one of the largest online grocery platforms. Walmart Grocery has over 100 million monthly active users who buy more than one hundred thousand products. We collect a portion of the dataset to test our model.
\end{itemize}

All three datasets contain the transactions about the interactions of users and items for each order. The items purchased in the same order are considered as a basket. As FENDER focuses on mitigating the long-term bias, we select the users with more than five baskets in the dataset. The statistics of the datasets are shown in Table \ref{tab:dataset}.

\begin{table}[]
    \centering
    \caption{The number of users, items, baskets, and average basket sizes in the datasets. }
    \resizebox{0.47\textwidth}{!}{\begin{tabular}{cccccc}
    \toprule
         Datasets & Users & Items & Baskets & Avg. basket size  \\
    \midrule
        Instacart & 56,305 & 49,689 & 479,164 & 8.51 \\
        Dunnhumby & 4,997 & 36,241 & 289,566 & 7.99 \\
        Walmart Grocery  & 61,365 & 89,839 & 538,171 & 8.77 \\
    \bottomrule
    
    \end{tabular}}
    
    \label{tab:dataset}
\end{table}

\subsubsection{Measurement metrics}
We utilize three metrics to measure the accuracy and diversity of FENDER's recommended list. Hit rate and NDCG are two widely used metrics in NBR. These two metrics measure FENDER's ability to learn the unbiased distribution improves accuracy. We also propose a novel negative Top-Frequency Ratio (nTFR), which measures the percentage of top-$k$ frequent items in the recommended list. Fewer frequent items in the recommended list mean there is more diversity. We define the nTFR as
\begin{equation}
    nTFR = 1 - \frac{N_{top\_fre}}{N_{rec\_item}},
\end{equation}
where $N_{top\_fre}$ and $N_{rec\_item}$ are the numbers of top frequent items and all items in the recommended list. In this paper, we choose $k$ as 20 based on the statistics of the datasets.

\subsubsection{Implementation}
The code is implemented by Tensorflow 1.12
and also compatible with Tensorflow version $>$ 2.0. The source codes and datasets are open-sourced\footnote{https://github.com/shawnlxh/FENDER}. We take the basket $t$ as testing set, basket $t-1$ as validation set and baskets $0$ to $t-2$ as training sets.

\subsection{Comparison Experiments}
This section compares our proposed FENDER with ten state-of-the-art baseline models. We measure our model on both accuracy and diversity. The detailed experiment results are shown in Table \ref{table:performance1}.

\subsubsection{Baseline models} We compare our method FENDER with ten state-of-the-art NBR algorithms. 

\begin{itemize}
    \item \textbf{PIF}. We rank the most frequently purchased items in all purchased items for the user and take them as the recommended items for the next basket.
    
    \item \textbf{BPR-MF}\cite{rendle2012bpr}. It is one of the most widely used baseline methods in recommender systems. We use all interactions of users and items as inputs to train the user and item embeddings.
    
    \item \textbf{propensity-MF}. It is an unbiased version of MF. In this model, we regard PIF as the propensity score and penalize it over different datapoints to derive an unbiased result.
    
    \item \textbf{DREAM}\cite{yu2016dynamic}. It is a Recurrent Neural Network (RNN)-based NBR model. DREAM captures the dynamic representations, as well as global sequential features, among baskets.
    
    \item \textbf{Triple2vec}\cite{wan2018representing}. It utilizes complementarity, compatibility, and loyalty to construct triplets for baskets and train user and item embeddings.
    
    \item \textbf{Beacon}\cite{le2019correlation}. It is a  hierarchical
    network to model basket sequences. It considers the relative importance of items and correlations among item pairs.
    
    \item \textbf{CLEA}\cite{qin2021world}. It extracts items relevant to the target item with contrastive learning to make denoising NBR.
    
    \item \textbf{Sets2Sets}\cite{hu2019sets2sets}. It is an NBR model based on the RNN and attention mechanism. A repetitive purchase pattern is also considered in this model.
    
    \item \textbf{MIT-GNN}\cite{liu2020basket}. It is the state-of-the-art GNN-based NBR recommendation model. Users, items, and baskets are nodes in the heterogeneous graph, and a translation-based model is utilized to learn the node embeddings. We take the PIF as the initial feature of the nodes.
    
    \item \textbf{TIFU-KNN}\cite{hu2020modeling}. It is the state-of-the-art PIF-based model in NBR. PIF is a crucial feature in the proposed KNN model to model the user's interest.
\end{itemize}

\begin{table*}[t]
  \centering
  \caption{Experiments on three datasets comparing our proposed FENDER framework with ten baseline models using the metrics: Hit Rate (HR),  Normalized Discounted Cumulative Gain (NDCG), and Negative Top-Frequency Ratio (nTFR). The bold and underlined numbers indicate the best and second-best results on each dataset and metric, respectively.}
    \resizebox{1.0\textwidth}{!}{\begin{tabular}{clccccccccccc}
    \toprule
    {Datasets}  
  &\multicolumn{3}{c}{Instacart}  &\multicolumn{3}{c}{Walmart Grocery} &\multicolumn{3}{c}{Dunnhumby}  \\
  \cmidrule(lr){1-1} \cmidrule(lr){2-4} \cmidrule(lr){5-7}  \cmidrule(lr){8-10} 
   {Models} & HR@20 & NDCG@20 & nTFR@20 & HR@20 & NDCG@20 & nTFR@20 & HR@20 & NDCG@20 & nTFR@20\\
    \midrule 
      PIF   &  0.863 & 0.416  & 0.136 & \underline{0.853} & 0.347 & 0.182 & 0.547 & 0.202 & 0.328 \\
      BPR-MF & 0.844 & 0.294  & \textbf{0.660} & 0.816 & 0.304 & 0.452 & \textbf{0.625} & \underline{0.330} & 0.416 \\
      propensity-MF   &  0.651 & 0.153  & 0.583 & 0.594 & 0.106 & \underline{0.454} & 0.267 & 0.206 & \underline{0.595} \\
      DREAM  &  0.705  & 0.116  & 0.574 & 0.621 & 0.153   & 0.437 & 0.248 & 0.102 & 0.485 \\
      Triple2Vec & 0.840 & 0.355 & 0.494 & 0.746  & 0.405 & 0.402 & 0.543 & 0.298 & 0.512   \\
      Beacon & 0.842 & 0.360 & 0.451 & 0.763  & 0.394 & 0.399 & 0.565 & 0.280 & 0.493   \\
      CLEA & 0.848 & 0.364 & 0.391 & 0.758  & 0.401 & 0.389 & 0.571 & 0.316 & 0.504   \\
      Sets2Sets   &  0.852 & 0.363  & 0.407 & 0.769 & 0.388 & 0.416 & 0.561 & 0.239 & 0.568 \\
      MIT-GNN  & 0.857 & 0.341 & 0.271 & 0.793 & 0.422 & 0.398 & 0.556 & 0.225 & 0.581 \\
      TIFU-KNN   &  \underline{0.885} & \underline{0.438}  & 0.161 & 0.847   & \underline{0.440} & 0.205 &   0.568 & 0.230  & 0.267 \\
      \textbf{FENDER} & \textbf{0.911} & \textbf{0.511}  & \underline{0.644} & \textbf{0.872} & \textbf{0.484} & \textbf{0.515} &  \underline{0.614} & \textbf{0.343}  & \textbf{0.679} \\
      \midrule
     Improvement & 2.94\% & 16.67\% & -2.4\% & 2.23\% & 10.00\% & 13.44\% & -1.76\% & 3.94\% & 14.12\% \\
    \bottomrule
    \end{tabular}}
  \label{table:performance1}
\end{table*}

\subsubsection{Experiment analysis}
Based on Table \ref{table:performance1}, we can compare the performance of all models on three datasets. Our observations of the experiment results are as follows:
\begin{itemize}
    \item In the comparison of recommendation accuracy, FENDER outperforms all baselines on all three datasets in NDCG@20. In comparing diversity, FENDER can defeat all baselines on the Walmart Grocery and Dunnhumby datasets and achieves the second-best in the Instacart dataset. At the same time, the best performing BPR-MF has very low NDCG@20. Therefore, we conclude that FENDER consistently outperforms all other baselines in accuracy and diversity on most datasets.
    Apart from FENDER, models considering frequency patterns in baskets (Sets2Sets, MIT-GNN, TIFU-KNN) achieve competitive accuracy compared to other baselines. However, the diversity of these models is relatively low, indicating the item frequency biases them.
    The unbiased model (propensity-MF) achieves high nTFR@20 but low HR@20 and NDCG@20, indicating that it tends to produce diverse results but loses its recommendation effectiveness. 
    
    \item We can also infer the distribution of the datasets from the table. In Instacart and Walmart Grocery, we can observe a high impact of PIF on recommendation results since the model PIF surpasses six other baselines on these datasets. This may be because the grocery scenario contains many perishable items that need to be purchased more frequently. However, the diversity of model PIF is the lowest, which means only focusing on PIF may deteriorate the diversity of the recommendation results. As for Dunnhumby, the performance of model PIF is relatively lower, indicating the PIF has a moderate impact on it.
    
    \item Compared with the second-best results on the three datasets we use, our model has a better performance on NDCG than on Hit rate. The average improvement on NDCG is 10.20\%, while on Hit Rate, it is only 1.14\%. It shows that FENDER can find a better ranking of items than other baseline models. The improvement in NDCG may result from the parameter $\omega$, which illustrates the inclination of PIF and confounders in the model training. In \hyperref[omega]{Section 5.5}, we have a more detailed analysis of this parameter.
\end{itemize}

\begin{figure}[tbp]
\centering
\subfigure[Instacart]{
\includegraphics[width=2in]{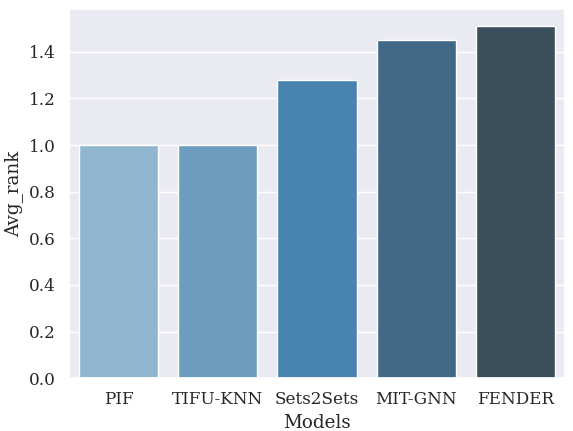}	
}
\quad
\subfigure[Walmart Grocery]{
\includegraphics[width=2in]{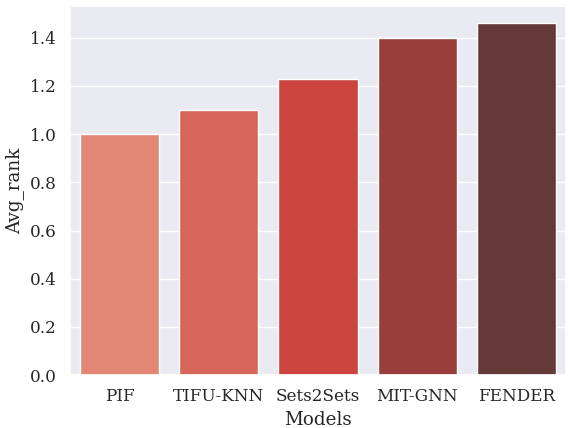}	
}
\caption{The average rank of the manually inserted items. We compare FENDER with four frequency-aware baseline models, i.e., PIF, TIFU-KNN, Sets2Sets, and MIT-GNN. }
\label{fig:avg_rank}
\end{figure}

\subsection{Robustness Experiment}
We conduct the robustness experiment by manually adding noise to the datasets. Specifically, we randomly choose an unrelated item for each user and insert it into all their purchased baskets. This way, we can make the unrelated items have the highest PIF compared to other items in the user's historical records. We conduct this experiment on Instacart and Walmart Grocery, as these datasets exhibit stronger repetitive purchase pattern and are easily influenced by PIF. To demonstrate the robustness of the models, we expect the model to identify the user's actual interest, and doesn't recommend the inserted item on the top ranking.

From the recommended lists of each model, we calculate the average rank of the inserted items among all users to represent the robustness of the model. A higher average rank of the inserted items means the model is more robust to the influence of PIF. As shown in Figure \ref{fig:avg_rank}, we observe that FENDER ranks the inserted items highest among the baselines on both datasets. Both Models PIF and TIFU-KNN  rank the inserted items near top-1, which are the lowest among all baseline models. This phenomenon indicates that FENDER can find some other features apart from PIF to generate recommendations with the deconfounder. Models PIF and TIFU-KNN rely on PIF as the main feature, so they rank the inserted items near the top of the recommended lists. Besides, models Sets2Sets and MIT-GNN all rank the inserted items higher than the top 1 because they consider other features (graph structure, item attention).

\begin{table*}[htbp]
\centering
  \caption{Case study of TIFU-KNN and FENDER. This table lists the top 10 recommended items for these two models. We re-index the item IDs in the dataset ranging from 1 to 45. The first five baskets are used as training data, and we list the next five predicted baskets for comparison.}
\makebox[0.6\textwidth][c]{
    \begin{tabular}{c|cccccccccc|cccccccccc}
\toprule
 Baskets& \multicolumn{10}{c|}{TIFU-KNN} & \multicolumn{10}{c}{FENDER} \\
\midrule
1 & 33  & 5 & 17 & 26  & 12 & 41  & 8 & 44 & 21 & 10
& 33  & 5 & 26 & 17  & 44 & 8  & 21 & 15 & 41  & 10 \\
2 & 33  & 5 & 17 & 41  & 26 & 12  & 44 & 8 & 15  & 7
& 33  & 5 & 17 & 44  & 8 & 26  & 19 & 7 & 15  & 41 \\
3 & 33  & 5 & 17 & 41  & 26 & 44  & 12 & 7 & 8  & 38
& 33  & 5 & 41 & 26  & 44 & 8  & 12 & 25 & 17  & 38 \\
4 & 33  & 5 & 17 & 41  & 26 & 44  & 12 & 8  & 15  & 7
& 33  & 5 & 17 & 26  & 8 & 27  & 44 & 20 & 19  & 12 \\
5 & 33  & 5  & 17 & 41  & 26 & 44  & 12 & 8 & 15  & 38
& 33  & 5 & 26 & 44  & 8 & 15  & 41 & 17 & 14  & 3 \\

\bottomrule
\end{tabular}%
}
\label{tab:case}%
\end{table*}%

\subsection{Case Study}
We conducted the case study on the Walmart Grocery dataset to compare the diversity of the recommended lists. We randomly select a user for this dataset and compare the recommended lists of TIFU-KNN \cite{hu2020modeling} and our proposed FENDER. TIFU-KNN is the most representative PIF-based model and performs well on the Walmart Grocery dataset according to Table \ref{table:performance1}. The comparison of these two models is listed in Table \ref{tab:case}. This table lists the top 10 items provided by the two models. The first five baskets are used as training data, and we compare the next five predicted baskets of the two models.

Based on Table \ref{tab:case}, we have the following observations:
\begin{itemize}
    \item The predicted lists of TIFU-KNN exhibit a homogenization pattern across different baskets, while FENDER has different patterns for different baskets. It demonstrates that our proposed FENDER can provide diverse recommendations for various visits of users. The diversity of recommendations improves user experience by creating a feeling of freshness.
    \item The number of distinct items in FENDER's recommended lists is 19, while TIFU-KNN only recommends 12 different items, 36.8\% less than FENDER. This indicates that FENDER has a broader browsing scope for items than TIFU-KNN. Broader browsing scope assists users in finding the potential items that they may be interested in, but have not purchased before.
    \item FENDER can still consider the repetitive purchasing pattern in the recommendation. According to Table \ref{tab:case}, items 33 and 5 always appear in the top 2 of the recommended lists. These two items are purchased almost every time by the users in the dataset. This shows that FENDER can find the items with PIF, while maintaining a unique shopping experience simultaneously.
\end{itemize}

\begin{figure}
\centering
    \includegraphics[height=1.9in]{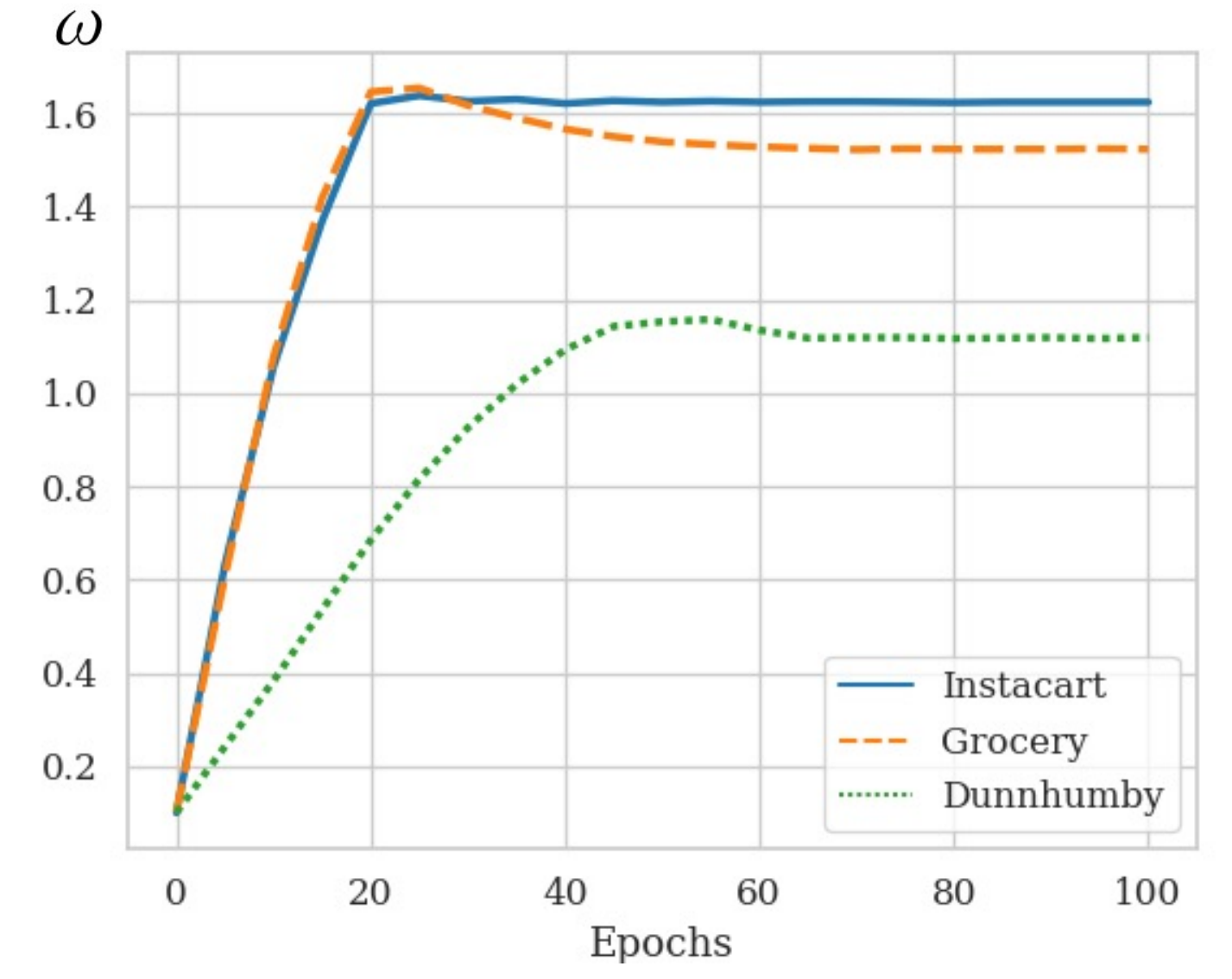}
    \caption{The variation trend of weight $\omega$ in datasets Instacart, Walmart Grocery, and Dunnhumby. We run 100 epochs to see the difference between the three curves.} 
    \label{fig:omega}
\end{figure}

\subsection{Parameter Analysis}\label{omega}
In this section, we investigate the selection of parameter $\omega$ in Eq. 6. $\omega$ illustrates the inclination of PIF and confounders in the model training. We demonstrate the variation trend during the training process for all three datasets. This experiment helps us understand the relations between confounders and PIF. Figure \ref{fig:omega} depicts parameter changes $\omega$ during 100 training epochs. We set the starting $\omega$ as 0.1 and the learning rate as 0.0001. 

Based on Figure \ref{fig:omega}, we observe that parameter $\omega$ converges to around 1.6 for datasets Instacart and Walmart Grocery, while on Dunnhumby, it converges to about 1.2. This phenomenon may result from the difference in the repetitive purchase pattern of the datasets. As described in Eq. 6, the allocation of PIF and confounders is $\omega$ and $(1-\omega)$. A higher $\omega$ denotes more differences between PIF and confounders as they are negatively correlated. According to Table \ref{table:performance1}, dataset Instacart and Walmart Grocery have a stronger repetitive purchase pattern than Dunnhumby, so the parameter $\omega$ is higher on Instacart and Walmart Grocery than on Dunnhumby.

%% file: text/5-RelatedWorks.tex
\section{Related Works}

\subsection{Next-Basket Recommendation}
Next Basket Recommendation~(NBR) predicts what the user will buy the next time (s)he visits the online platform.  Previous works mainly focus on two aspects: 1) modeling the complementary pattern among historical basket  records.~\cite{mcauley2015inferring,wan2018representing,xu2019modeling}; and 2) considering the frequency pattern for the purchased items \cite{hu2020modeling, faggioli2020recency, hu2019sets2sets}. Complementary pattern is the complementary relationship between items and we can leverage it to know which items will be added next based on the existing items in the basket. Sceptre~\cite{mcauley2015inferring} is proposed to model and predict relationships between items from the text of their reviews and the corresponding descriptions.  DREAM~\cite{yu2016dynamic} proposes to use RNN structure to exploit the dynamics of user embeddings to complete BR. Triple2vec~\cite{wan2018representing} improves within-basket recommendation via (user, item A, item B) triplets sampled from baskets.  MIT-GNN~\cite{liu2020basket} uses heterogeneous graph embeddings to complete BR. All these works simply aggregate the information within the basket, while having no discussion regarding the complete feedback loop. Le et al. \cite{le2019correlation} propose a hierarchical network to model the basket sequences based on the correlation of items in each basket. CLEA \cite{qin2021world} extracts items relevant to the
target item for NBR with a Gumbel Softmax-based denoising generator. Wang et al. devise Intention2Basket~\cite{wang2020intention2basket} including three modules: Intention Recognizer, Coupled Intention Chain Net, and Dynamic Basket Planner to model the heterogeneous intentions behind baskets. All the models above learn from the complementary pattern of baskets to model users in different innovative ways, but none of them consider the frequency pattern behind items.

Apart from the complementary pattern, the frequency pattern is also demonstrated to be useful in the NBR problem. Hu et al. \cite{hu2020modeling} leverage Personalized Item Frequency (PIF) as a key feature in the KNN model and achieved the state-of-the-art performance. Faggioli et al. \cite{faggioli2020recency} propose a recency-aware user-wise popularity in the collaborative filtering method to model the repetitiveness and loyalty of the user. Sets2Sets \cite{hu2019sets2sets} is an end-to-end learning approach based on an encoder-decoder framework to predict the subsequent sets of elements. Sets2Sets also considers the repeated elements pattern in the model to improve performance. The frequency or repetition-based models perform well in the offline setting, but this kind of strategy is detrimental to the users' experience in the long term. If a model relies on frequency pattern heavily, the scope of the item set in the recommendation list will be significantly narrowed. In this paper, we want to find a balance to increase the diversity of the items in the recommendation list while maintaining a satisfactory performance from the frequency pattern.

\subsection{Recommender System Debiasing}
Recommender systems have achieved wide success in different tasks \cite{liu2020basket, li2020dynamic, li2021pre}, but most of them are still suffering from the biases in the recommendation results. There is rich literature dealing with biases in recommender systems. While a few studies can take a holistic view of the debiasing task on recommender systems, the vast majority of studies are organized either from the probabilistic perspective or the system design perspective.
From the probabilistic perspective, scientists try to analyze the distribution of the recommendation dataset or results.
The item exposure is regarded as a crucial factor in the recommendation data generating process and is modeled explicitly in many studies~\cite{liang2016causal2,liang2016modeling,sato2020unbiased,wang2020causal}. 
Many studies believe that the bias in recommender systems is induced by the \textit{missing-not-at-random} nature of the dataset, and focus on the following outcome of this selection bias~\cite{christakopoulou2020deconfounding,wang2020information,saito2020unbiased,saito2020asymmetric,liu2021mitigating}. 
From the system perspective, scientists address the biases resulting from the flaw of the system. 
Some studies discover that the recommendation result is biased toward a particular item or user group and try to address such as popularity bias~\cite{abdollahpouri2017controlling,abdollahpouri2019managing,zhang2021causal,zhu2021popularity,li2021user} or conformity bias~\cite{zheng2021disentangling}.
\cite{schnabel2016recommendations,yang2018unbiased,gruson2019offline} introduce debiased estimators to correct the selection bias in the recommendation evaluation step.
\cite{liu2020general,wang2021combating,zeng2021causal} leverage a small volume of unbiased uniform data to calibrate the biased dataset. 
Furthermore, the studies of feedback loops regard the recommender system and users as a whole system and hope the recommender system can achieve a balance during its interactions with users~\cite{chaney2018algorithmic,xu2020adversarial,wang2021deconfounded}. There are more related works regarding other biases, such as position bias in learning-to-rank systems~\cite{joachims2017unbiased,agarwal2018counterfactual,ovaisi2020correcting}.

Although the recommender system debiasing problem attracts much attention in academia and industry, this topic is still challenging due to the critical generality/variance tradeoff: if a method can handle multiple biases with minor modification, its variance must be high, making itself difficult to train. Moreover, because of the different sources of the biases, there is no panacea to deal with this problem in recommender systems. In different scenarios, the biases can come from different aspects, such as item popularity or unobserved confounders. In this paper, we discuss the frequency bias and devise a model only for NBR. We first propose a causal model to explain the cause of such bias. Then, we propose a flexible and scalable method to reduce the bias systematically.

%% file: text/6-Conclusion.tex
\section{Conclusion \& Future Works}
This paper discusses the frequency bias issue in the NBR problem. Frequency bias results from the RS model relying heavily on the PIF feature. In the bias amplification loop, the frequency bias accumulates and increases the recommendation results' homogenization.
Frequency bias gradually narrows a user's browsing scope on the online platform and eventually deteriorates the user's experience. 
We propose FENDER, a model based on the deconfounder architecture, to deal with frequency bias. FENDER factorizes the PIF into latent embeddings to eliminate the frequency bias. 
Experimental results show that FENDER can improve the accuracy of the recommendation results and the diversity of the recommended items.

From a method perspective, our model can be viewed as a variation of sequential deconfounder \cite{hatt2021sequential,bica2020time} and is a sequential model. While conventional sequential model/deconfounder requires many mediate variables to learn the hidden state of the next timestep, our model only needs a pre-calculated feature (PIF) to summarize users' interests and thus speeds up the calculation process significantly. In the future, similar ideas can be used to design models. We can select representative features to substitute certain modules in complex machine learning models to improve the efficiency and effectiveness of models simultaneously. 

\section{Acknowledgement}
This work is supported in part by NSF under grants III-1763325, III-1909323, III-2106758, and SaTC-1930941.